\documentstyle[aps,pra]{revtex}
\title{Unambiguous Discrimination Between Linearly-Independent Quantum States}
\author{Anthony Chefles \\
       Department of Physics and Applied Physics, University of Strathclyde,
       Glasgow G4 0NG, Scotland \\
       e-mail: tony@phys.strath.ac.uk \\\mbox{}\\}  

\input amssym.def
\input epsf
\epsfverbosetrue

\def\id{\hat{\leavevmode\hbox{\small1\kern-3.8pt\normalsize1}}}%
\begin{document}
\maketitle
\thanks{PACS: 03.65.Bz}

\begin{abstract}

The theory of generalised measurements is used to examine the problem
of discriminating unambiguously between non-orthogonal pure quantum
states. Measurements of this type never give erroneous results,
although, in general, there will be a non-zero probability of a result
being  inconclusive.  It is shown that only linearly-independent states
can be unambiguously discriminated.  In addition to examining the
general properties of such measurements, we discuss their application
to entanglement concentration.

\end{abstract}

\newpage

\section{Introduction}
\renewcommand{\theequation}{1.\arabic{equation}}
\setcounter{equation}{0}

In classical mechanics, it is possible in principle to determine the state of any physical system. The state of a classical system is described by its canonical coordinates and momenta, which can be measured simultaneously to arbitrary precision.  In quantum mechanics, however, the state of a system is represented as a vector in a Hilbert space, and is not itself an observable quantity.  Precise determination of a completely unknown state vector is precluded by the nature of the quantum measurement process.  It is only when the state belongs to a known orthogonal set that it can be infallibly determined, by a standard von Neumann measurement.  

When confronted with the problem of trying to discriminate between non-orthogonal states, we must accept that no strategy will correctly reveal the state of the system with unit probability.  It is then of considerable theoretical and practical importance to find the optimum strategy, that is, the one with the highest probability of giving the correct result. This may be regarded as the central problem of {\em quantum detection theory}, which was  established in the 1960s and 70s through the seminal work of Helstrom[1] and Holevo[2].  An indispensable tool here is the theory of generalised measurements or {\em quantum operations}[3], which lays down the necessary and, in principle, sufficient conditions for the physical realisability of quantum state transformations.   The optimum detector problem consists of finding the quantum operation which gives the minimum average error probability for a known ensemble of states whose {\em a priori} probabilities are also given. 

More recently, it has been realised that another, and in some circumstances, potentially more useful approach to discriminating between non-orthogonal states is possible.  Ivanovic[4] demonstrated the existence of a detector which is able to discriminate {\em unambiguously} between a pair of non-orthogonal pure states, that is, with zero error probability.  This measurement does not contravene the laws of quantum mechanics since, for non-orthogonal states, it has less than unit probability of giving an answer at all.  There will be a non-zero probability that the measurement returns an inconclusive result.  The Ivanovic measurement has subsequently been streamlined and investigated in greater depth by Dieks[5] and Peres[6].  The Ivanovic-Dieks-Peres (IDP) measurement is a generalised measurement involving a unitary interaction with a two-state ancilla particle.  A von Neumann measurement is then performed on the latter.  This has two outcomes.  One outcome maps both states of the particle of interest onto orthogonal states, which can be unambiguously discriminated.  The other outcome causes them to be mapped onto the same state, giving an inconclusive result and completely erasing the information we wish to obtain.  The IDP measurement gives the lowest average probability of obtaining inconclusive results for unambiguous measurements when both states, which we denote by $|{\psi}_{\pm}>$, have equal {\em a priori} probabilities of 1/2.  The probability of the IDP measurement resulting in an inconclusive result is
\begin{equation}
P_{IDP}=|<{\psi}_{+}|{\psi}_{-}>|.
\end{equation}

Measurements of this kind have been performed in the laboratory by Huttner {\em et al}[7].  Weak pulses of light were prepared in non-orthogonal polarisation states, a fraction of which were converted into orthogonal ones by a loss mechanism.  Applications of this type of measurement to eavesdropping on quantum cryptosystems have been discussed by Ekert {\em et al}[8], and some of the consequences of carrying them out locally on part of an entangled system are discussed in [9].  It is, in fact, possible to attain a lower probability of inconclusive results than the $P_{IDP}$ limit provided we are prepared to allow for errors[10].  Given a fixed probability of inconclusive results $P_{I}{\le}P_{IDP}$, the minimum error probability $P_{E}$ saturates the inequality
\begin{equation}
P_{E}P_{D}{\ge}\frac{1}{4}(P_{IDP}-P_{I})^{2}
\end{equation}
where $P_{D}=1-P_{I}-P_{E}$ is the successful discrimination probability.  Setting $P_{I}=0$ gives the well-known Helstrom bound[1] $P_{E}{\ge}(1-[1-|<{\psi}_{+}|{\psi}_{-}>|^{2}]^{1/2})/2$, which corresponds to the highest probability of giving the correct result among all possible strategies.

The IDP analysis has recently been extended by Jaeger and Shimony[11] to the more general case of two states with unequal {\em a priori} probabilities.  While this analysis exhausts the theoretical problem of optimising unambiguous state-discrimination measurements for a pair of pure states, little or no attention has been given to the general problem of error-free discrimination between multiple quantum states.  This paper is concerned with the general theory of unambiguous state-discrimination.  In section II, it is shown that the necessary and, by construction,  sufficient condition for the existence of unambiguous measurement strategies for a given set of  non-orthogonal pure states is that the states must be linearly-independent, and we explore the general properties of the appropriate measurement transformations.   These measurements are found not to be of the standard von Neumann type, but are more general quantum operations.  We then  describe the variational problem for optimum measurements.  In section III, we show that the results of Jaeger and Shimony follow from our general formalism when restricted to two states, and derive the minimum probability of inconclusive results for $n$ states among a class of constrained measurements, the constraint being that all states have equal conditional probabilities of being identified.  The optimum such measurement is used in section IV, where we extend our earlier discussion[9] of the relationship between the IDP measurement and entanglement concentration[12] to the $n$-state case.  Here, we describe a method based on state-discrimination which converts a fraction of an ensemble of $n$-level imperfectly-entangled systems into maximally-entangled ones. 

\section{Unambiguous State Discrimination}
\renewcommand{\theequation}{2.\arabic{equation}}
\setcounter{equation}{0}

Consider a quantum system known to be in one of the non-orthogonal
pure states $|{\psi}_{j}>$, where $j=1,..,n$. We denote by ${\cal H}$
the Hilbert space spanned by these states.  We wish to design a quantum
measurement whose result tells us which state the system was prepared in,
and furthermore, which never gives errors.  The non-orthogonality of the
$|{\psi}_{j}>$ means that no standard von Neumann measurement on ${\cal H}$ can
fulfil these requirements, and we must look to generalised measurements for the solution to the problem.  Any generalised measurement can be
expressed in terms of a set of linear transformation operators ${\hat A}_{k}$ acting on ${\cal H}$[3].
The only constraint on these operators is that they must satisfy the resolution
of the identity

\begin{equation}
\sum_{k}{\hat A}^{\dagger}_{k}{\hat A}_{k}={\id}.
\end{equation}

Each ${\hat A}_{k}$ corresponds to a distinct outcome of the operation.
Taking the system to be prepared with the initial density operator ${\hat
\rho}$, the probability $P_{k}$ of the $k$th outcome is ${\mathrm
Tr{\hat \rho}}{\hat A}^{\dagger}_{k}{\hat A}_{k}$.  Correspondingly, the density operator
is transformed according to ${\hat \rho}{\rightarrow}{\hat A}_{k}{\hat
\rho}{\hat A}^{\dagger}_{k}/P_{k}$.  Here, we are concerned with quantum operations which allow us to unambiguously discriminate between
the $n$ states  $|{\psi}_{j}>$, and see that such a measurement must have $n$
outcomes signalling the detection of each of the states.  We assume, for the sake of simplicity, that only one outcome of the operation corresponds to the detection of each of the states, as this affects none of the conclusions that follow.   An additional outcome is however required to allow for inconclusive results,
the possibility of which is unavoidable if the measurement strategy is
required to be error-free, as has been found in studies of the problem of
discriminating between just two states[4-11].  Thus, the state-discrimination
measurement will be characterised by the $n$ operators ${\hat A}_{k}$
corresponding to the states and a further transformation operator ${\hat
A}_{I}$ generating inconclusive results.  These operators must satisfy

\begin{equation}
{\hat A}^{\dagger}_{I}{\hat A}_{I}+\sum_{k}{\hat A}^{\dagger}_{k}{\hat 
A}_{k}={\id}.
\end{equation}

The $k$th outcome should only arise if the initial state is  
$|{\psi}_{k}>$, implying the following constraint:

\begin{equation}
<{\psi}_{j}|{\hat A}^{\dagger}_{k}{\hat A}_{k}|{\psi}_{j}>=P_{j}{\delta}_{jk}.
\end{equation}

We now proceed to show that this zero-errors condition can only be met if 
the $|{\psi}_{j}>$ are linearly-independent.  To accomplish this, let us attempt to express them as superpositions of each other by writing 
\begin{equation}
|{\psi}_{j}>=\sum_{r}f_{jr}|{\psi}_{r}>,
\end{equation}
and determine the constraints imposed upon the $f_{jr}$ by Eq. (2.3).  Inserting this expression for $|{\psi}_{j}>$ into Eq. (2.3), we find
\begin{equation}
\sum_{r,r'}f^{*}_{jr'}f_{jr}<{\psi}_{r'}|{\hat A}^{\dagger}_{k}{\hat 
A}_{k}|{\psi}_{r}>=P_{j}{\delta}_{jk}.
\end{equation}
The last factor on the left of Eq. (2.5) is easily simplified with the aid of the 
Cauchy-Schwarz inequality
\begin{equation}
|<{\psi}_{r'}|{\hat A}^{\dagger}_{k}{\hat 
A}_{k}|{\psi}_{r}>|^{2}{\le}<{\psi}_{r}|{\hat A}^{\dagger}_{k}{\hat 
A}_{k}|{\psi}_{r}><{\psi}_{r'}|{\hat A}^{\dagger}_{k}{\hat A}_{k}|{\psi}_{r'}>
\end{equation}
and Eq. (2.3), which together imply that 
\begin{equation}
<{\psi}_{r'}|{\hat A}^{\dagger}_{k}{\hat 
A}_{k}|{\psi}_{r}>=P_{r}{\delta}_{rr'}{\delta}_{rk}.
\end{equation}

Substitution of this expression into Eq. (2.5) then gives 
$|f_{jr}|^{2}={\delta}_{jr}$, which implies that none of the states can be a linear superposition of the others.  They must be linearly-independent.  

We can easily establish the form of the ${\hat A}_{k}$ using the constraint 
Eq. (2.7).  One useful consequence of Eq. (2.7) is that the states 
$A_{k}|{\psi}_{k}>$ are orthogonal, so that we may write
\begin{equation}
{\hat A}_{k}|{\psi}_{k}>=P^{1/2}_{k}|{\phi}_{k}>
\end{equation}
where the $|{\phi}_{k}>$ form an orthonormal basis for ${\cal H}$.  A further 
useful property of ${\hat A}_{k}$ which follows from Eq. (2.7) is that it 
annihilates the subspace spanned by all $|{\psi}_{j}>$ for $j{\neq}k$.  Let us 
denote this subspace as ${\cal H}_{k}$, whose one-dimensional orthogonal 
complement we take to be spanned by the state $|{\psi}^{\perp}_{k}>$.  From these two observations, we can infer 
that ${\hat A}_{k}$ has the form
\begin{equation}
{\hat A}_{k}=\frac{P^{1/2}_{k}}{<{\psi}^{\perp}_{k}|{\psi}_{k}>}|{\phi}_{k}><{\psi}^{\perp}_{k}|.
\end{equation} 

It is worth remarking that the states $|{\psi}^{\perp}_{k}>$ are also linearly-independent.  If there was a superposition $\sum_{k}a_{k}|{\psi}^{\perp}_{k}>$ which was equal to the zero vector, then acting upon this state with the operator $\sum_{k}|{\psi}_{k}><{\psi}_{k}|$ would give $\sum_{k}a_{k}<{\psi}_{k}|{\psi}^{\perp}_{k}>|{\psi}_{k}>=0$, which is impossible since the $|{\psi}_{k}>$ are linearly-independent.  Note that the $<{\psi}_{k}|{\psi}^{\perp}_{k}>$ factors are all non-zero due to the linear-independence of the $|{\psi}_{k}>$.  

We have assumed that the states $|{\psi}_{k}>$ span the Hilbert space upon which the transformation operators act.  If, however, we examine a situation where the Hilbert space of the system is larger than the space ${\cal H}$ spanned by the $|{\psi}_{k}>$, a more general form of the ${\hat A}_{k}$ is permitted.  We briefly discuss these more general measurements in the Appendix, and show that there always exists a measurement restricted to ${\cal H}$ which is optimal, in the sense that it maximises the average success probability.

Let us examine the problem of maximising the average success probability.  We denote by ${\eta}_{j}$ the {\em a priori} probability that the system was prepared in the state $|{\psi}_{j}>$, whereupon we find that the probability of successfully determining the state of the system is
\begin{equation}
P_{D}(|{\psi}_{j}>, {\eta}_{j})=\sum_{j}{\eta}_{j}<{\psi}_{j}|{\hat 
A}^{\dagger}_{j}{\hat A}_{j}|{\psi}_{j}>=\sum_{j}{\eta}_{j}P_{j}.
\end{equation}
Thus, we interpret $P_{j}$ as the conditional probability that, given the system 
was prepared in the state $|{\psi}_{j}>$, this state will be identified.

This probability is constrained by the positivity 
of the operator ${\hat A}^{\dagger}_{I}{\hat A}_{I}$ which,  together with the 
decomposition of unity in Eq. (2.2), is equivalent to saying that none of the  
eigenvalues of the positive operator ${\sum}_{j}{\hat A}^{\dagger}_{j}{\hat 
A}_{j}$ are greater than 1.  Thus, we can state the variational problem whose solution is the maximum value of $P_{D}(|{\psi}_{j}>, {\eta}_{j})$ in the following way.  Let us define the probability operator
\begin{equation}
\hat{\Pi}(\{P_{j}\})=\sum_{j}{\hat A}^{\dagger}_{j}{\hat 
A}_{j}=\sum_{j}q_{j}P_{j}|{\psi}^{\perp}_{j}><{\psi}^{\perp}_{j}|,
\end{equation}
where we have let
\begin{equation}
q_{j}=|<{\psi}^{\perp}_{j}|{\psi}_{j}>|^{-2}.
\end{equation}
We also denote by ${\lambda}(\{P_{j}\})$ the maximum eigenvalue of ${\hat {\Pi}}(\{P_{j}\})$.  Then we wish to vary the $P_{j}$ so as to obtain the maximum value of $P_{D}(|{\psi}_{j}>, {\eta}_{j})$ subject to the constraint that ${\lambda}(\{P_{j}\}){\le}1$.  

We can prove that the average discrimination probability is equal to 1 only when the states $|{\psi}_{j}>$ are orthogonal.  If $P_{D}(|{\psi}_{j}>, {\eta}_{j})=1$, then it is clear from Eq. (2.10) that the conditional discrimination probabilities $P_{j}$ must also be 1.  Assuming that this is the case, let us consider taking the expectation value of $\hat{\Pi}(\{P_{j}=1\})$ for an arbitrary quantum state $|{\psi}>$ and determine the conditions under which its maximum attainable value is 1.  Due to the linear-independence of the $|{\psi}_{j}>$ and the fact that they span the Hilbert space, we can expand any state as a unique superposition of them,
\begin{equation}
|{\psi}>=N^{-1/2}\sum_{j}c_{j}|{\psi}_{j}>,
\end{equation}
such that $\sum_{j}|c_{j}|^{2}=1$ and where the normalisation factor $N$ is
\begin{equation}
N=\sum_{j,j'}c^{*}_{j'}c_{j}<{\psi}_{j'}|{\psi}_{j}>.
\end{equation}
Using the observation that $<{\psi}^{\perp}_{j'}|{\psi}_{j}>$ is non-zero only when $j=j'$, we find that
\begin{equation}
<{\psi}|\hat{\Pi}(\{P_{j}=1\})|{\psi}>=N^{-1}.
\end{equation}
Therefore, we seek the conditions under which the normalisation factor $N$ is not less than 1 for any state.  This can be accomplished by defining the Hermitian operator
\begin{equation}
{\hat W}=\sum_{j,j'}<{\psi}_{j'}|{\psi}_{j}>|w_{j'}><w_{j}|,
\end{equation}
where the $|w_{j}>$ form an orthonormal basis for the Hilbert space.  Corresponding to every state $|{\psi}>$ given by Eq. (2.13), there is a unique state $|\tilde{\psi}>=\sum_{j}c_{j}|w_{j}>$, and the expectation value of  ${\hat W}$ for this state is $N$.  This allows us to show that the necessary and sufficient condition for $N{\ge}1$ is that the $|{\psi}_{j}>$ must be orthogonal, by the following argument.  The trace of ${\hat W}$ is equal to $n$, the dimension of the Hilbert space, implying that its average eigenvalue is 1.  Therefore, if the eigenvalues are unequal, some must be less than unity, as must $N$ for some states $|\tilde{\psi}>$.  This can be averted only if all of the eigenvalues are equal to 1, implying that ${\hat W}$ is identity operator, which clearly, from Eq. (2.16) would require the $|{\psi}_{j}>$ to be orthonormal.  Therefore, unambiguous state-discrimination measurements have unit probability of succeeding only when the states are orthogonal.   

The maximum value of $P_{D}(|{\psi}_{j}>, {\eta}_{j})$ will always be obtained for $P_{j}$ such that ${\lambda}(\{P_{j}\})=1$.  Were this not the case, we could multiply all $P_{j}$ by an equal factor bringing ${\lambda}(\{P_{j}\})$ up to 1, with a corresponding increase in the value of $P_{D}(|{\psi}_{j}>, {\eta}_{j})$.  One important consequence of this is that, for optimum measurements, when inconclusive results are obtained, the set of states $|{\psi}_{j}>$ is mapped onto a linearly-dependent set.  The proof of this is simple: given that $\hat{\Pi}(\{P_{j}\})$ has an eigenvalue 1,  the subspace spanned by the corresponding eigenvectors is, as a consequence of 
Eq. (2.2), the kernel of ${\hat A}^{\dagger}_{I}{\hat A}_{I}$.  Consider a 
state $|{\phi}>$ lying in this kernel.  We can expand this state as 
$|{\phi}>=\sum_{j}a_{j}|{\psi}_{j}>$.  Normalisation is unimportant here.  If 
we also write ${\hat A}_{I}|{\psi}_{j}>=|{\psi}'_{j}>$, then  
\begin{equation}
\sum_{j}a_{j}|{\psi}'_{j}>=0.
\end{equation}
Thus, the (unnormalised) states $|{\psi}'_{j}>$ are linearly-dependent.  From our proof that only linearly-independent states can be unambiguously discriminated, we see that for the maximum value of $P_{D}(|{\psi}_{j}>, {\eta}_{j})$, when an inconclusive result is obtained, one cannot make a further attempt to determine the state of the system unless one is prepared to use a strategy which has a non-zero error probability.  In the particular case of two states, they are mapped onto the same state, rendering any kind of further attempt impossible. 

The general solution to the variational problem for arbitrary states $|{\psi}_{j}>$ and {\em a priori} probabilities ${\eta}_{j}$ is presently unknown.  It is doubtful that a closed form can be found for the general maximum value of the average discrimination probability $P_{D}(|{\psi}_{j}>, {\eta}_{j})$.  However, certain special cases can be treated analytically.  One is the problem for just two states, for which the variational problem can be solved exactly.   In the following section we show that the results of Jaeger and Shimony[11] can be derived from our general formalism.  We also determine the optimum value of $P_{D}(|{\psi}_{j}>, {\eta}_{j})$ in the case of $n$ states with the additional constraint that all $P_{j}$ are equal. 

\section{Examples}
\renewcommand{\theequation}{3.\arabic{equation}}
\setcounter{equation}{0}

To find the maximum value of $P_{D}(|{\psi}_{j}>, {\eta}_{j})$  for a pair of non-orthogonal pure states, it is convenient to denote them by $|{\psi}_{\pm}>$, and to exploit the fact that they can always be represented as
\begin{equation}
|{\psi}_{\pm}>={\cos}{\theta}|+>{\pm}{\sin}{\theta}|->,
\end{equation}
where the angle ${\theta}$ lies in the range $[0,{\pi}/4]$ and the states $|{\pm}>$ constitute an orthogonal basis for the space spanned by $|{\psi}_{\pm}>$.  The system may be represented as a spin-1/2 
particle, and $|{\pm}>$ taken to be the eigenstates of ${\hat {\sigma}}_{z}$ 
with eigenvalues ${\pm}1$.  We find that  
\begin{equation}
|{\psi}^{\perp}_{\pm}>={\sin}{\theta}|+>{\pm}{\cos}{\theta}|->,
\end{equation}
where $<{\psi}^{\perp}_{+}|{\psi}_{-}>=<{\psi}^{\perp}_{-}|{\psi}_{+}>=0$ and  $<{\psi}^{\perp} _{\pm}|{\psi}_{\pm}>={\sin}2{\theta}$.  Given that the states $|{\psi}_{\pm}>$ have respective {\em a priori} probabilities ${\eta}_{\pm}$, we wish to find the maximum value of the discrimination probability $P_{D}(|{\psi}_{j}>, {\eta}_{j})$ given by Eq. (2.10), subject to the constraint that the maximum eigenvalue of the operator 
\begin{equation}
\hat{\Pi}(\{P_{j}\})=\frac{1}{2{\sin}^{2}{2\theta}}\{(P_{+}+P_{-})(1-{\hat \sigma}_{z}{\cos}2{\theta})+(P_{+}-P_{-}){\hat \sigma}_{x}{\sin}2{\theta}\}
\end{equation}
is equal to 1.  The eigenvalues of this operator are easily determined, and we find that the greater of them is equal to 1 only when
\begin{equation}
(1-P_{+})(1-P_{-})=|<{\psi}_{+}|{\psi}_{-}>|^{2},
\end{equation}
where we have used ${\cos}2{\theta}=|<{\psi}_{+}|{\psi}_{-}>|$.  Eq. (3.4) is  precisely the constraint obtained by Jaeger and Shimony[11] when considering unambiguous state-discrimination using a unitary interaction with a two-state ancilla followed by a von Neumann measurement.  The remainder of their analysis follows from maximising the average discrimination probability with respect to the constraint in Eq. (3.4) for $0{\le}P_{\pm}{\le}1$.  Given that it is possible, without loss of generality, to take ${\eta}_{+}{\ge}{\eta}_{-}$, the Jaeger-Shimony minimum probability $P_{JS}$ of an inconclusive result is
\begin{equation}
P_{JS}=2({\eta}_{+}{\eta}_{-})^{1/2} |<{\psi}_{+}|{\psi}_{-}>|
\end{equation}
when $({\eta}_{-}/{\eta}_{+})^{1/2}{\ge}|<{\psi}_{+}|{\psi}_{-}>|$ and
\begin{equation}
P_{JS}={\eta}_{+}|<{\psi}_{+}|{\psi}_{-}>|^{2}+{\eta}_{-}
\end{equation}
whenever $({\eta}_{-}/{\eta}_{+})^{1/2}{\le}|<{\psi}_{+}|{\psi}_{-}>|$.  In this latter case, the optimum measurement is of the von Neumann type where ${\hat A}_{+}=|{\psi}^{\perp}_{+}><{\psi}^{\perp}_{+}|, {\hat A}_{I}=|{\psi}_{-}><{\psi}_{-}|$ and ${\hat A}_{-}=0$.  We see that the state $|{\psi}_{-}>$ is never detected.  Thus, for sufficiently large values of the overlap $|<{\psi}_{+}|{\psi}_{-}>|$ or the {\em a priori} probablity ${\eta}_{+}$ of the $|{\psi}_{+}>$ state, the optimum unambiguous measurement will consist of detecting only this state, and this occurs when the previously unknown state of the system is projected onto $|{\psi}^{\perp}_{+}>$, the state orthogonal to $|{\psi}_{-}>$. 
                                                                                
One further special case which is exactly soluble is that of $n$ states subject to the constraint that all $P_{j}$ are equal to some constant $P$.  It follows from Eq. (2.10) that the average discrimination probability $P_{D}(|{\psi}_{j}>, {\eta}_{j})$ is also equal to $P$, whose value can be deduced from our knowledge that the largest eigenvalue of $\hat{\Pi}(\{P_{j}=P)\}$ must be 1.  This simply implies that 
\begin{equation}
P_{D}(|{\psi}_{j}>, {\eta}_{j})=P=({\mathrm max}\sum_{j}q_{j}|<{\psi}|{\psi}^{\perp}_{j}>|^{2})^{-1},
\end{equation}
that is, the optimal discrimination probability is simply the reciprocal of the maximum eigenvalue of the operator $\sum_{j}q_{j}|{\psi}^{\perp}_{j}><{\psi}^{\perp}_{j}|$.  We shall make use of this measurement in the next section, where we show how state-discrimination measurements, when performed locally on part of a larger, imperfectly entangled system, can leave the system in a maximally-entangled state.

\section{State Discrimination and Entanglement Concentration}
\renewcommand{\theequation}{4.\arabic{equation}}
\setcounter{equation}{0}

In this section, we examine the effect of carrying out state-discrimination measurements on a subsystem which is part of a larger system in an imperfectly entangled pure state.  We generalise to the case of $n$-level systems our earlier demonstration[9] that the IDP measurement, which discriminates between a pair of non-orthogonal states, can serve as a basis for {\em entanglement concentration}, that is, can convert a fraction of a number of imperfectly entangled two-state systems into maximally-entangled ones such as singlets.  In this special case, our method is equivalent to the `Procrustean' technique proposed by Bennett {\em et al}[12].

Let us begin by considering a pair of $n$-level quantum systems.  Our analysis can easily be generalised to systems with Hilbert spaces of different dimension.  Giving our subsystems the labels 1 and 2, the state space ${\cal H}$ of the composite is the tensor product ${\cal H}_{1}{\otimes}{\cal H}_{2}$ of their individual Hilbert spaces.  Any pure state $|{\phi}>$ in ${\cal H}$ can be written as a Schmidt decomposition[13]
\begin{equation}
|{\phi}>=\sum_{j=0}^{n-1}c_{j}|{\alpha}_{j}>{\otimes}|{\beta}_{j}>,
\end{equation}
that is, as a single sum where the $|{\alpha}_{j}>$ and $|{\beta}_{j}>$ are orthonormal bases for ${\cal H}_{1}$ and ${\cal H}_{2}$ respectively.  Note that in this section, we have adopted the convention of letting the index $j$ run from 0 to $n-1$, which turns out to be more convenient in what follows.   The state is normalised so that $\sum_{j}|c_{j}|^{2}=1$.  We assume that all of the $c_{j}$ are non-zero.  Maximally-entangled states are those for which all of the $|c_{j}|^{2}$ are equal to $1/n$.  Let us now imagine that we have a large number of composite systems prepared in the state $|{\phi}>$, and show how a modified state-discrimination measurement of the type described in sections II and III, performed locally on one part of each entangled pair, can transform a fraction of these systems into maximally-entangled states.

To do this, we must express the state $|{\phi}>$ in a more appropriate form.  Corresponding to the basis $|{\alpha}_{j}>$ for ${\cal H}_{1}$, we can define a conjugate basis
\begin{equation}
|{\gamma}_{k}>=\frac{1}{\sqrt n}\sum_{j}{\exp}\left[{\frac{-2{\pi}ijk}{n}}\right]|{\alpha}_{j}>.
\end{equation}
The basis $|{\gamma}_{k}>$ is canonically conjugate to the $|{\alpha}_{j}>$ in the sense discussed by Kraus[14] and Pegg {\em et al}[15].  The orthonormality of the $|{\gamma}_{k}>$ is a consequence of the identity
\begin{equation}
\frac{1}{n}\sum_{j}{\exp}\left[{\frac{-2{\pi}ij(k-k')}{n}}\right]={\delta}_{kk'}.
\end{equation}
It follows that the state $|{\phi}>$ can be rewritten as
\begin{equation}
|{\phi}>=\frac{1}{{\sqrt n}}\sum_{k}|{\gamma}_{k}>{\otimes}|{\psi}_{k}>
\end{equation}
where the new states $|{\psi}_{k}>$ for the second system are given by
\begin{equation}
|{\psi}_{k}>=\sum_{j}c_{j}{\exp}\left[{\frac{2{\pi}ijk}{n}}\right]|{\beta}_{j}>.
\end{equation}
These states are normalised, although they are not orthogonal; if they were, $|{\phi}>$ would be maximally-entangled.  They are, however, linearly-independent.  To show this, let us look at the consquences of supposing that 
\begin{equation}
\sum_{k}a_{k}|{\psi}_{k}>=0
\end{equation}
for some coefficients $a_{k}$.  Substituting into this sum the expansion in Eq. (4.5), we find that the condition
\begin{equation}
\sum_{k}a_{k}{\exp}\left[{\frac{2{\pi}ijk}{n}}\right]=0
\end{equation}
must be satisfied for all $j$.  Multiplication of Eq. (4.7) throughout by ${\mathrm e}^{\frac{-2{\pi}ijk'}{n}}$, summing over $j$ and making further use of Eq. (4.3) gives $a_{k'}=0$.  Thus, the only way that Eq. (4.6) can be true is if all coefficients in this sum are zero, so the $|{\psi}_{k}>$ are linearly-independent.   

In Eq. (4.4), $|{\phi}>$ has equal coefficients, so it is the non-orthogonality of the $|{\psi}_{k}>$ that prevents this state from being maximally-entangled.  However, recall that a measurement which unambiguously discriminates between these states does so by mapping them onto orthogonal states.  Having established their linear-independence, we see that such an operation is possible for  $|{\phi}>$.

The state-discrimination measurement of section II must, however, be adapted somewhat to achieve our present aim; the transformation operators ${\hat A}_{k}$ defined in Eq. (2.9) would convert $|{\phi}>$ into a mixed state.  A way to avoid this comes from viewing the state-discrimination measurement as a two-stage procedure.  First, the states $|{\psi}_{k}>$ are mapped onto the orthogonal states $|{\phi}_{k}>$.  A von Neumann measurement is then carried out which projects onto this latter basis.  Entanglement concentration requires only the first stage of this operation to be carried out, so let us define the orthogonalisation operator
\begin{equation}
{\hat A}_{O}=\sum_{k}{\hat A}_{k}=\sum_{k}\frac{P^{1/2}_{k}}{<{\psi}^{\perp}_{k}|{\psi}_{k}>}|{\phi}_{k}><{\psi}^{\perp}_{k}|.
\end{equation}
This operator performs the required mapping of the $|{\psi}_{k}>$ onto the orthogonal $|{\phi}_{k}>$.  That state-discrimination can be regarded as this operation followed by a von-Neumann measurement is easily seen from the fact that ${\hat A}_{k}=|{\phi}_{k}><{\phi}_{k}|{\hat A}_{O}$.  Note that we also have
\begin{equation}
{\hat A}^{\dagger}_{O}{\hat A}_{O}=\sum_{k}{\hat A}^{\dagger}_{k}{\hat 
A}_{k}={\id}-{\hat A}^{\dagger}_{I}{\hat A}_{I},
\end{equation}
where ${\hat A}_{I}$ is the operator introduced in Eq. (2.2).  In state-discrimination attempts, this operator generates inconclusive results.  In the present context of entanglement concentration, its action results in a failure to produce the desired maximally-entangled state. 

The state given by ${\hat A}_{O}|{\phi}>$, when normalised, is a standard maximally-entangled state when the $P_{k}$ in Eq. (4.8) are set to the same value, $P$.  The probability $P_{C}$ of successful conversion of $|{\phi}>$ into a maximally-entangled state is 
\begin{equation}
P_{C}=<{\phi}|{\hat A}^{\dagger}_{O}{\hat A}_{O}|{\phi}>=P,
\end{equation}
that is, the probability of transforming $|{\phi}>$ into a maximally-entangled state using our method is simply the probability of successfully discriminating between the states $|{\psi}_{k}>$ when we require the conditional discrimination probabilites of these states to be equal.  We gave the formal solution, Eq. (3.7), to the problem of finding the maximum value of $P$, and hence $P_{C}$ in the preceding section.  However,  it turns out that for $|{\psi}_{k}>$ given by Eq. (4.5), a particularly simple expression for $P$ can be obtained.  We wish to evaluate the reciprocal of the maximum eigenvalue of the operator $\sum_{k}q_{k}|{\psi}^{\perp}_{k}><{\psi}^{\perp}_{k}|$.   The states $|{\psi}^{\perp}_{j}>$ are found to be
\begin{equation}
|{\psi}^{\perp}_{k}>=N^{-1/2}\sum_{j}c^{*-1}_{j}{\exp}\left[{\frac{2{\pi}ijk}{n}}\right]|{\beta}_{j}>,
\end{equation}
where the normalisation factor $N=\sum_{j}|c_{j}|^{-2}$ and the $q_{j}$, which we introduced in Eq. (2.13), are
\begin{equation}
q_{j}=\frac{N}{n^{2}}.
\end{equation}
Collecting these results together, we find, with the aid of Eq. (4.3), that
\begin{equation}
\sum_{k}q_{k}|{\psi}^{\perp}_{k}><{\psi}^{\perp}_{k}|=\frac{1}{n}\sum_{j}|c_{j}|^{-2}|{\beta}_{j}><{\beta}_{j}|.
\end{equation}

Since the $|{\beta}_{j}>$ are orthonormal, the eigenvalues of this operator are simply read off to be $1/n|c_{j}|^{2}$.  Thus, the maximum probability of converting $|{\psi}>$ into a maximally-entangled state using our concentration prodedure is 
\begin{equation}
P_{C}=n{\times}{\mathrm min}(|c_{j}|^{2}).
\end{equation}

It is interesting to see what happens when the entanglement concentration attempt fails.  This naturally involves examining the operator ${\hat A}_{I}$, which is constrained by ${\hat A}_{O}$ and Eq. (4.9) to have the form  ${\hat A}_{I}={\hat U}[{\id}-{\hat A}^{\dagger}_{O}{\hat A}_{O}]^{1/2}$, 
where ${\hat U}$ is any unitary operator acting on ${\cal H}_{2}$.  This operator annihilates all terms $|{\alpha}_{j}>{\otimes}|{\beta}_{j}>$ in the Schmidt decomposition Eq. (4.1) which correspond to the smallest of the $|c_{j}|^{2}$.  Therefore, a failure to produce the desired maximally entangled state will remove the possibility of the subsystems 1 and 2 being found in the states $|{\alpha}_{j}>$ and $|{\beta}_{j}>$, and thus their correlation.  In the case of just two states, the initial entangled state becomes a product state.

\section{Summary}

In this paper, we have considered the problem of unambiguous state-discrimination.  Our particular aims were to determine the general conditions under which this is possible, and the form of the appropriate measurement operators.   As we have seen, one can discriminate among a set of pure states $|{\psi}_{j}>$ if and only if they are linearly-independent.  The measurement itself consists of mapping these states onto an orthogonal set $|{\phi}_{j}>$, followed by a von Neumann measurement in this basis.  

Only orthogonal states can be unambiguously discriminated with unit probability.    In general, there will be a non-zero probability of obtaining inconclusive results.  It is clearly of interest to study the conditions under which the average discrimination probability $P_{D}(|{\psi}_{j}>, {\eta}_{j})$ reaches its maximum value for a given set of states $|{\psi}_{j}>$ with {\em a priori} probabilities ${\eta}_{j}$.  We have formulated the appropriate variational problem, and shown that its solution for just two states is the same as that obtained by Jaeger and Shimony[11] using a specific measurement model.  We have also found the maximum discrimination probability for $n$ states with the additional constraint that all $P_{j}$ are equal.  Clearly, however, the general problem deserves further study.  An important goal of future work is the determination of the maximum discrimination probability for cases other than those considered here.  A still more challenging problem is to devise practical techniques for accurately preparing a given quantum system in one of many non-orthogonal pure states, and realising the appropriate transformations. 

As an application of the general formalism of sections II and III, we showed in section IV how state-discrimination may be related to entanglement concentration.  It is well known that any inseperable pure state with two subsystems can be expressed as a biorthogonal expansion known as the Schmidt decomposition, Eq. (4.1).  If the state is not maximally-entangled, the coefficients are not equal, but can be made so if we choose to express one of the subsystems in terms of non-orthogonal, but nevertheless linearly-independent states.  The first stage of the state-discrimination operation, namely the mapping of these states onto an orthogonal basis, can then be used to produce a maximally-entangled state, by acting only on one of the subsystems.  However, as with state-discrimination, this process is not guaranteed to be successful, and we run the risk of degrading rather than enhancing the correlations when it fails.

\section*{Appendix}
\renewcommand{\theequation}{A.\arabic{equation}}
\setcounter{equation}{0}

Here, we give a brief discussion of more general unambiguous state-discrimination measurements where the Hilbert space ${\cal H}$ spanned by the $|{\psi}_{j}>$ is a proper subspace of the space spanned by the possible states of the system.  We shall denote this larger space by ${\cal H}_{S}$.  This is the direct product ${\cal H}_{S}={\cal H}{\oplus}{\cal H}^{'}$, where ${\cal H}^{'}$ is the orthogonal complement of ${\cal H}$ within ${\cal H}_{S}$.  A generalisation of the measurement described in section II is possible under these circumstances since there exists more than one state in ${\cal H}_{S}$ which has a non-zero overlap with the state $|{\psi}_{j}>$ and is orthogonal to all $|{\psi}_{k}>$ for $k{\neq}j$.   For measurements restricted to ${\cal H}$, these states are uniquely, up to a phase, given by $|{\psi}^{\perp}_{j}>$.  However, in the more general case we are considering here, any state of the form
\begin{equation}
|{\psi}^{\perp}_{Sj}>={\mu}_{j}|{\psi}^{\perp}_{j}>+{\nu}_{j}|{\chi}_{j}>
\end{equation}
may be used, where the $|{\chi}_{j}>$ are arbitrary normalised states in ${\cal H}^{'}$ and $|{\mu}_{j}|^{2}+|{\nu}_{j}|^{2}=1$.  The subscript $S$ will be used in what follows for states and operators in the enlarged space ${\cal H}_{S}$.  Clearly, 
\begin{equation}
<{\psi}_{k}|{\chi}_{j}>=<{\psi}^{\perp}_{k}|{\chi}_{j}>=0.
\end{equation}

The discussion in section II leading to the construction of the transformation operators ${\hat A}_{j}$, given by Eq. (2.9), can easily be repeated in this broader setting.  We find that the corresponding generalised transformation operators ${\hat A}_{Sj}$ have the form 
\begin{equation}
{\hat A}_{Sj}=\frac{P^{1/2}_{j}}{<{\psi}^{\perp}_{Sj}|{\psi}_{j}>}|{\phi}_{Sj}><{\psi}^{\perp}_{Sj}|.
\end{equation} 
The $|{\phi}_{Sj}>$ can be any $n$ orthonormal states in ${\cal H}_{S}$.  As is the case with measurements restricted to ${\cal H}$, the sum $\sum_{j}{\hat A}^{\dagger}_{Sj}{\hat A}_{Sj}$ does not equal the identity operator, so we are obliged to introduce an operator ${\hat A}_{SI}$ generating inconclusive results which satisfies
\begin{equation}
{\hat A}^{\dagger}_{SI}{\hat A}_{SI}+\sum_{j}{\hat A}^{\dagger}_{Sj}{\hat 
A}_{Sj}={\id}_{S},
\end{equation}
where ${\id}_{S}$ is, of course, the identity on ${\cal H}_{S}$.  The probabilities $P_{j}$ are restricted by the constraint that the maximum eigenvalue ${\lambda}_{S}(\{P_{j}\})$ of the probability operator
\begin{equation}
\hat{\Pi}_{S}(\{P_{j}\})=\sum_{j}{\hat A}^{\dagger}_{Sj}{\hat 
A}_{Sj}=\sum_{j}q_{j}P_{j}|{\psi}^{\perp}_{Sj}><{\psi}^{\perp}_{Sj}|
\end{equation}
must not exceed 1.  For optimum measurements, giving the maximum value of $P_{D}(|{\psi}_{j}>, {\eta}_{j})$, we find that ${\lambda}_{S}(\{P_{j}\})=1$.  This can be established by an argument similar to that given in section II which showed that ${\lambda}(\{P_{j}\})=1$ for optimum measurements confined to ${\cal H}$ 

We shall now prove that, for given states $|{\psi}_{j}>$ and {\em a priori} probabilities ${\eta}_{j}$, the maximum value of the average discrimination probability $P_{D}(|{\psi}_{j}>, {\eta}_{j})$ can always be obtained by considering measurements which are restricted to the space ${\cal H}$.  We begin with the observation that, for any values of the probabilities $P_{j}$ and any state $|{\psi}>$ in ${\cal H}$, we have
\begin{equation}
<{\psi}|\hat{\Pi}(\{P_{j}\})|{\psi}>=<{\psi}|\hat{\Pi}_{S}(\{P_{j}\})|{\psi}>.
\end{equation}

The support of $\hat{\Pi}(\{P_{j}\})$ is ${\cal H}$, so we are entitled to apply Eq. (A.6) when $|{\psi}>$ is an eigenstate of $\hat{\Pi}(\{P_{j}\})$ corresponding to its maximum eigenvalue ${\lambda}(\{P_{j}\})$.  If $|{\psi}>$ is also an eigenstate of $\hat{\Pi}_{S}(\{P_{j}\})$ with corresponding maximum eigenvalue ${\lambda}_{S}(\{P_{j}\})$, then both of these extremal eigenvalues are seen from Eq. (A.6) to be equal.  If not, however, then (A.6) implies that ${\lambda}_{S}(\{P_{j}\})$ must exceed ${\lambda}(\{P_{j}\})$, giving the general inequality  
\begin{equation}
{\lambda}_{S}(\{P_{j}\}){\ge}{\lambda}(\{P_{j}\})
\end{equation}
We can use this result to show how the assumption that the maximum value of $P_{D}(|{\psi}_{j}>, {\eta}_{j})$ cannot be attained for measurements restricted to ${\cal H}$ leads to a contradiction.  If the $P_{j}$ correspond to the optimum measurement, giving the maximum value of $P_{D}(|{\psi}_{j}>, {\eta}_{j})$ in Eq. (2.10), and that this measurement cannot be restricted to ${\cal H}$, then ${\lambda}_{S}(\{P_{j}\})=1$.  This implies that a restricted measurement with the same probabilities $P_{j}$ would give ${\lambda}(\{P_{j}\}){\le}1$.  Thus, we can consider another restricted measurement with probabilities $P^{'}_{j}=P_{j}/{\lambda}(\{P_{j}\})$ such that $P^{'}_{j}{\ge}P_{j}$, giving at least as high a value for the average discrimination probability.  The maximum eigenvalue ${\lambda}(\{P^{'}_{j}\})$ of the corresponding probability operator ${\hat {\Pi}}(\{P^{'}_{j}\})$ is 1, so this measurement is allowed.  Therefore, for any measurement on the larger Hilbert space ${\cal H}_{S}$, we can construct a measurement restricted to ${\cal H}$ which is at least as good, contradicting the premise.

\section*{Acknowlegements}

I would like to thank Prof. Stephen M. Barnett and Dr. John Jeffers for
valuable discussions, and gratefully acknowledge financial support by
the UK Engineering and Physical Sciences Research Council (EPSRC).

\newpage

\section*{References}

\mbox{}\\{[1] C. W. Helstrom, {\em Quantum 
Detection and Estimation Theory}, (Academic Press, New York,
1976).}\\\mbox{}\\{[2] A. S. Holevo, {\em J. Multivar. Anal.} {\bf 3} (1973) 337.}\\\mbox{}\\{[3]
K. Kraus, {\em States, Effects and Operations}, (Number 190 in
Lecture Notes in Physics, Springer, Berlin, 1983).}\\\mbox{}\\{[4]
I. D. Ivanovic, {\it Phys. Lett. A} {\bf 123} (1987) 257.}\\\mbox{}\\{[5]
D. Dieks, {\em Phys. Lett. A} {\bf 126} (1988) 303.}\\\mbox{}\\{[6]
A. Peres, {\it Phys. Lett. A} {\bf 128} (1988) 19.}\\\mbox{}\\{[7] B. Huttner, A. Muller J. D. Gautier, H. Zbinden
and N.  Gisin, {\em Phys. Rev A} {\bf 54} (1996) 3783.}\\\mbox{}\\{[8]
A. Ekert, B. Huttner, G. M. Palma and A. Peres, {\it Phys. Rev. A} {\bf 50} (1994) 1047.}\\\mbox{}\\{[9]  A. Chefles and S. M. Barnett, {\em Phys. Lett. A} {\bf 236} (1997) 177.}\\\mbox{}\\{[10] A. Chefles and
S. M. Barnett, {\em J. Mod. Opt.} (in press).}\\\mbox{}\\{[11] G. Jaeger and
A. Shimony, {\em Phys. Lett. A} {\bf 197} (1995) 83.}\\\mbox{}\\{[12] C. H. Bennett, H. J. Bernstein, S. Popescu and
B. Schumacher, {\it Phys. Rev. A} {\bf 53} (1996) 2046.}\\\mbox{}\\{[13]
A. Ekert and P. L. Knight, {\em Am. J. Phys.} {\bf 63} (1995) 415.} \\\mbox{}\\{[14]
K. Kraus, {\em Phys. Rev. D} {\bf 35} (1987) 3070.}\\\mbox{}\\{[15]
D. T. Pegg, J. A. Vaccaro and S. M. Barnett, {\em J. Mod. Opt.} {\bf 37}
(1990) 1704.}
\end{document}